\documentclass[aps,pra,twocolumn,superscriptaddress,longbibliography,footnote]{revtex4-1}
\usepackage{bm}
\usepackage{amsmath,mathtools}
\usepackage{amssymb}
\usepackage{amsthm}
\usepackage{amsfonts}
\usepackage{graphicx}
\usepackage{epsfig}
\usepackage{epstopdf}
\usepackage[hidelinks,colorlinks=true,linkcolor=blue,citecolor=blue,anchorcolor=blue,filecolor=blue,menucolor=blue,runcolor=blue,urlcolor=blue]{hyperref}
\usepackage{color}
\usepackage{setspace}

\begin{document}


\title{Scattering Problem in Bose-Einstein Condensates with Magnetic Domain Wall}

\author{Mei Zhao}
\email{These authors contribute equally to this paper.}
\affiliation{Institute of Modern Physics, Northwest University, Xi'an, 710127, China}
\author{Lijia Jiang}
\email{These authors contribute equally to this paper.}
\affiliation{Physics department, Northwest University, Xi'an, 710127, China}
\affiliation{Shaanxi Key Laboratory for Theoretical Physics Frontiers, Xi'an 710127, China}
\affiliation{Peng Huanwu Center for Fundamental Theory, Xi'an 710127, China}
\affiliation{Fundamental Discipline Research Center for  Quantum Science and technology of Shaanxi Province, Xi'an 710127, China}
\author{Tao Yang}
\email{yangt@nwu.edu.cn}
\affiliation{Institute of Modern Physics, Northwest University, Xi'an, 710127, China}
\affiliation{Shaanxi Key Laboratory for Theoretical Physics Frontiers, Xi'an 710127, China}
\affiliation{Peng Huanwu Center for Fundamental Theory, Xi'an 710127, China}
\affiliation{Fundamental Discipline Research Center for  Quantum Science and technology of Shaanxi Province, Xi'an 710127, China}
\author{Jun-Hui Zheng}
\email{junhui.zheng@nwu.edu.cn}
\affiliation{Institute of Modern Physics, Northwest University, Xi'an, 710127, China}
\affiliation{Shaanxi Key Laboratory for Theoretical Physics Frontiers, Xi'an 710127, China}
\affiliation{Peng Huanwu Center for Fundamental Theory, Xi'an 710127, China}
\affiliation{Fundamental Discipline Research Center for  Quantum Science and technology of Shaanxi Province, Xi'an 710127, China}

\begin{abstract}
We present a comprehensive theoretical study of linear wave scattering from magnetic domain walls with varied twist angles \(\Theta\) in spin-\(1/2\) Bose-Einstein condensates (BECs). Using a gauge transformation, we show that scattering observables depend solely on the total twist \(\Theta\), independent of chirality. Within the Bogoliubov-de Gennes (BdG) framework, we develop a transfer-matrix method to compute reflection and transmission coefficients for incident phonons and free particles. Our results reveal a scattering threshold at the Zeeman energy \(E = \hbar\Omega_0\), separating a pure phonon regime from multi-channel scattering involving both collective and single-particle excitations above threshold. Above a critical twist angle \(\Theta_c\), the effective spin rotation deviates from the imposed twist angle, leading to comb-like density modulations and Fano-like resonances below the threshold $\hbar\Omega_0$. The transition probability between phonon and particle channels is strongly tunable with \(\Theta < \Theta_c\), enhanced for odd multiples of \(\pi\) but suppressed for even multiples. These findings establish twist-engineered domain walls as a versatile platform for controlling quantum transport, with implications for atomtronic devices and quantum simulation.
\end{abstract}

\maketitle
\section{Introduction}

The exploration of topological defects and nonlinear coherent structures in quantum fluids \cite{mermin1979topological} represents a vibrant frontier at the intersection of condensed matter physics \cite{nagaosa2013topological,parkin2008}, atomic physics \cite{kawaguchi2010topological,Borgh2012,borgh2014imprinting,Matthews1999,Choi2012}, and quantum optics \cite{hallacy2025nonlinear}. Among these systems, BECs provide a uniquely pristine and controllable platform for investigating such phenomena. The rich internal degrees of freedom in spinor condensates \cite{kawaguchi2012spinor}, in particular, enable the emulation of diverse effects from conventional condensed matter systems, including magnetism \cite{Black2007}, superconductivity \cite{Bighin2025}, and spin transport \cite{McGuirk2010,Cao2025,Watabe2012,Jimenez2019,Kudo2013}. Within this context, the study of topological defects such as solitons \cite{Dum1998,kumar2010collision,Qu2016}, vortices \cite{fetter2009rotating}, and domain walls \cite{Watabe2012,jin2011spin,Coen2001} has yielded profound insights into non-equilibrium dynamics, spontaneous symmetry breaking, and quantum coherence \cite{Madison2000,sadler2006spontaneous,Saito2007}. Moreover, the exquisite experimental control over interactions, geometry, and external fields in ultracold atomic gases opens promising avenues for applications in quantum information processing and atomtronics, where engineered defects can serve as conduits or barriers for atomic and spin currents \cite{atala2014observation,Scott2003}.


Domain walls, serving as interfaces that separate distinct magnetic \cite{malozemoff2013magnetic,emori2013current} or superfluid phases \cite{kuratsuji2021domain,Li2010,Son2002,Zhang2005,chen2023vector}, constitute a canonical class of topological defects. In spinor BECs, these structures manifest as smooth, spatially varying textures in the condensate's spin polarization. They are typically stabilized by an external magnetic field with a spatially dependent direction, imposing a preferred spin orientation that rotates across the cloud. The fundamental types include the N\'{e}el wall, where spin rotates within a plane containing the wall axis, and the Bloch wall, where rotation occurs in a plane perpendicular to this axis \cite{malozemoff2013magnetic,kittel2018introduction}. 
Understanding the static and dynamic properties of these domain walls including their stability and interactions with other excitations, is essential for advancing spin manipulation and transport in quantum gases \cite{Watabe2012,zhu2025magnetic,Yu2021,Hurst2019,HolandaRibeiro2023}.

A crucial aspect of this endeavor involves characterizing how quantum fluctuations and collective excitations scatter from these topological structures or impurities ~\cite{ohashi2021perfect,alotaibi2019scattering,gaul2008anisotropic,Watabe2015,Watabe2011,Watabe2011a,Watabe2011b}, which
constitutes a fundamental probe of a system's linear response and reveals information about its underlying order. In BECs, this is formally described within the BdG formalism, which provides a mean-field framework for studying the spectrum of small-amplitude excitations atop a nontrivial ground state~\cite{dalfovo1999theory,leggett2001bose}. When the ground state itself hosts a topological defect like a domain wall, the scattering problem becomes inherently multi-component and is influenced by synthetic gauge fields arising from the spatially varying spin texture~\cite{Li2010,Watabe2012,zhu2025magnetic,dalibard2011colloquium}.

Despite these advances, a systematic understanding of how the twist angle of a magnetic domain wall governs scattering processes remains incomplete. 
Moreover, the role of spin texture in mediating transitions between collective (phonon) and single-particle excitations warrants detailed investigation. In this work, we present a comprehensive theoretical study of linear wave scattering from magnetic domain walls with systematically varied twist angles in spinor BECs. We begin by deriving the mean-field ground state in the presence of a position-dependent Zeeman field that creates domain walls with total twist angles $\Theta$. We demonstrate the fundamental equivalence of different chiral configurations through a gauge and a global transformation, establishing that physical observables depend solely on the total twist angle rather than specific chirality. To address the scattering problem, we linearize the dynamics around the ground state, obtaining the BdG equations.  We systematically construct the scattering matrix via a transfer matrix method, enabling computation of reflection and transmission coefficients for incident phonons and free particles.

Our results elucidate how the embedded spin texture governs scattering processes, revealing pronounced threshold effects at the Zeeman splitting $E = \hbar \Omega_0$ and demonstrating that the effective spin rotation can differ substantially from the imposed twist angle in order to minimize the total energy. Such inconsistencies generate comb-like density structures leading to Fano-like resonances \cite{Miroshnichenko2010,Vicencio2007,yasir2016controlled,Zheng2018}. Furthermore, we establish that the transition probability between collective and single-particle modes can be strongly manipulated by the twist angle, with significant transitions occurring for $\Theta =\pi$ but suppressed for $\Theta =2\pi$.


\section{System and Hamiltonian}

The dynamics of a two-component BEC are governed by two coupled Gross-Pitaevskii equations (GPEs). For a quasi-one-dimensional system with SU(2)-symmetric contact interactions ($g_{\uparrow\uparrow} = g_{\downarrow\downarrow} = g_{\uparrow\downarrow} = g$) and a position-dependent Zeeman field $\boldsymbol{\Omega}(x)$, the GPEs read
\begin{equation}
i\hbar\frac{\partial}{\partial t}
\begin{pmatrix}
\psi_\uparrow \\
\psi_\downarrow
\end{pmatrix} = H_{\text{lab}}
\begin{pmatrix}
\psi_\uparrow \\
\psi_\downarrow
\end{pmatrix},
\end{equation}
where the effective Hamiltonian is
\begin{equation}
H_{\text{lab}} = -\frac{\hbar^2}{2m} \partial_x^2  - \frac{\hbar}{2} \boldsymbol{\Omega}(x) \cdot \hat{\boldsymbol{\sigma}} + g n(x) \mathbb{I}_2.
\end{equation}
Here, $\hat{\boldsymbol{\sigma}} = (\hat{\sigma}_x, \hat{\sigma}_y, \hat{\sigma}_z)$ denotes the Pauli matrices, $ \mathbb{I}_2$ is the $2\times 2$ identity matrix, $n(x) = |\psi_\uparrow|^2 + |\psi_\downarrow|^2$ is the total density, and the mean-field interaction $g n(x)$ acts identically on both spin components. In the presence of a Zeeman field,
\begin{equation}
\boldsymbol{\Omega}(x) = \Omega_0 \left( \cos\alpha \sin\theta(x),\ \sin\alpha \sin\theta(x),\ \cos\theta(x) \right),
\end{equation}
where $\theta(x) = \Theta \cdot [1 + \tanh (x/W)]/2$ introduces a twist angle $\Theta$, the system supports a magnetic domain wall with a width of order $W$. The chirality $\alpha$ determines the plane in which the spin rotates: $\alpha = 0$ yields a N\'eel wall, while $\alpha = \pi/2$ gives a Bloch wall. Notably, the twist angle $\Theta$ can be any real value and is not restricted to $\pi$.

We perform a position-dependent gauge transformation
\begin{equation}
U_\alpha(x) = \exp\left[ - i \frac{\theta(x)}{2} (-\sin\alpha\, \sigma_x + \cos\alpha\, \sigma_y) \right]
\end{equation}
that aligns Zeeman fields along the $+\hat{z}$ direction \cite{zheng2020theory}. In this rotated frame, $\Psi_{\text{new}}(x) = U_\alpha^\dagger(x) \Psi_{\text{lab}}(x)$, and the Hamiltonian becomes
\begin{equation}
H_{\text{new}}^{\alpha} =U_\alpha^\dagger(x)  H_{\text{lab}}  U_\alpha(x) = -\frac{\hbar^2}{2m} D_x^2 - \frac{\hbar\Omega_0}{2}\sigma_z + gn(x)\mathbb{I}_2,
\end{equation}
where
$
D_x = \partial_x  - i{\theta'(x)}(-\sin\alpha\, \sigma_x + \cos\alpha\, \sigma_y)/{2}
$
is the covariant derivative that encodes spin-orbit coupling arising from the spin texture. The Hamiltonian with different chiralities are all equivalent via the global rotation
$
R(\alpha) = \exp\left( -i {\alpha}\sigma_z/{2} \right)
$
that satisfies the transformation $R^\dagger(\alpha) (-\sin\alpha\, \sigma_x + \cos\alpha\, \sigma_y) R(\alpha) = \sigma_y$:
\begin{equation}
 H_0 \equiv R^\dagger(\alpha) H_{\text{new}}^{\alpha} R(\alpha) = H_{\text{new}}^{\alpha=0}.
\end{equation}
For $\alpha =0$, $D_x = \partial_x - i{\theta'(x)}\sigma_y/{2}$. Consequently, domain walls of different chiralities represent the same physical system, which depends solely on the twist $\Theta$ and the width $W$, and is independent of $\alpha$.

\section{Scattering problem}

Without loss of generality we henceforth focus on $\alpha =0$ and treat the GPEs in the new spin frame:
\begin{equation}
i\hbar\frac{\partial}{\partial t} \begin{pmatrix} \psi_\uparrow \\ \psi_\downarrow \end{pmatrix} = H_0\begin{pmatrix} \psi_\uparrow \\ \psi_\downarrow \end{pmatrix}
.
\end{equation}
The ground state $\Psi_0 =\begin{pmatrix} \psi_{\uparrow,0} \\ \psi_{\downarrow,0} \end{pmatrix}$ satisfies
$
 H_0 \Psi_0  = \mu \Psi_0$,
where the Hamiltonian is self-consistently evaluated with the ground-state density $n_0(x) = |\psi_{\uparrow,0}|^2 + |\psi_{\downarrow,0}|^2$.
This state is obtained by imaginary-time evolution of the GPEs. The system's total energy consists of three parts:
\begin{equation}\label{en}
  E_{\text{tot}} = E_{\text{kin}} + E_{\text{Zeeman}} + E_{\text{int}},
\end{equation}
where $E_{\text{kin}} = ({\hbar^2}/{2m}) \int dx \, (D_x \Psi_0)^\dagger(D_x \Psi_0)$ is the kinetic energy,  $E_{\text{Zeeman}} = -({\hbar\Omega_0}/{2}) \int dx \, \big(|\psi_{\uparrow,0}|^2 - |\psi_{\downarrow,0}|^2\big)$ is the Zeeman energy, and $E_{\text{int}} = ({g}/{2}) \int dx \, n_0^2$ is the interaction energy.
Below we study the scattering problem of linear waves on this ground-state background.
We write the wave function as
\begin{equation}
\begin{pmatrix} \psi_\uparrow(x,t) \\ \psi_\downarrow(x,t) \end{pmatrix}
= e^{-i\mu t/\hbar} \left[ \Psi_0(x) + \delta\Psi(x,t) \right].
\end{equation}
We consider the fluctuation for a single normal mode of energy  $E =\hbar\omega$:
\begin{equation}
\delta\Psi(x,t) = U(x) e^{-i\omega t} + V^*(x) e^{i\omega t},
\end{equation}
with components
$
U(x) = \begin{pmatrix} u_\uparrow(x) \\ u_\downarrow(x) \end{pmatrix}$ and
$V(x) = \begin{pmatrix} v_\uparrow(x) \\ v_\downarrow(x) \end{pmatrix}
$.
Linearising GPEs in $\delta\Psi$ and $\delta\Psi^*$ yields the BdG equations:
\begin{equation}\label{bdg2}
E \begin{pmatrix} U(x) \\ V(x) \end{pmatrix}
=
\mathcal{H}_{\text{BdG}}
\begin{pmatrix} U(x) \\ V(x) \end{pmatrix},
\end{equation}
where $ \mathcal{H}_{\text{BdG}} =  \tau_z \mathcal{H}'_{\text{BdG}} $ with $\mathcal{H}'_{\text{BdG}}$ being Hermitian and $\tau_z = \begin{pmatrix} \mathbb{I}_2 & 0 \\ 0 & -\mathbb{I}_2 \end{pmatrix}$. $\mathcal{H}'_{\text{BdG}} = \mathcal{T} + \mathcal{V}$ contains kinetic energy and effective potential,
\begin{equation}
\mathcal{T} = -\frac{\hbar^2}{2m} \begin{pmatrix}
D_x^2 & 0 \\
0 & D_x^2
\end{pmatrix}, \quad
\mathcal{V} = \begin{pmatrix}
V  & V_{\times} \\
V_{\times}^* & V^*
\end{pmatrix},
\end{equation}
where $
V = -{\hbar\Omega_0}\sigma_z/{2} + [gn_0(x)- \mu]\mathbb{I}_2 + g\Psi_0\Psi_0^\dagger $ and $ V_{\times} =  g\Psi_0\Psi_0^T $.

%
%
%

\subsection{Away from the Domain Wall}
In the region far away from the domain wall, the gauge field vanishes, i.e., $\theta'(x) \to 0$. There the ground state is a uniform BEC occupying only the spin-up component, $\Psi_0(x) = (\sqrt{n_0}, 0 )^T$, and the chemical potential is $\mu = g n_0 -\hbar \Omega_0/2$. The BdG Hamiltonian becomes
\begin{equation}\label{bdg0}
\mathcal{H}_{\text{BdG}} \rightarrow
\begin{pmatrix}
\hat{T} + g n_0 & 0 & g n_0 & 0 \\
0 & \hat{T} + \hbar\Omega_0 & 0 & 0 \\
-g n_0 & 0 & - \hat{T} - g n_0 & 0 \\
0 & 0 & 0 & -\hat{T} - \hbar\Omega_0
\end{pmatrix},\end{equation}
with $\hat T=   -({\hbar^2}/{2m})\partial^2_x $. For a given energy $E$
there are eight solutions for the BdG equation in this region.

The Bogoliubov spectrum for the spin-up block from Eq.\eqref{bdg0} is
$
E = \sqrt{\epsilon_k^2 + 2g n_0 \epsilon_k}$ with $\epsilon_k = {\hbar^2 k^2}/{2m}
$.
This quartic equation in $ k $ yields four roots $ \pm k_1 $  and $ \pm k_2 $:
\begin{eqnarray}
k_1 &=& \sqrt{2m/\hbar^2}  \sqrt{-g n_0 + \sqrt{E^2 + (g n_0)^2}}, \\
k_2 &=&  i \sqrt{2m/\hbar^2} \sqrt{g n_0 + \sqrt{E^2 + (g n_0)^2}},
\end{eqnarray}
and the associated eigenvectors are
\begin{equation}
\phi^{1,2}(x)
= \begin{pmatrix} U_{k_1} \\ 0 \\ V_{k_1} \\0 \end{pmatrix} e^{\pm i k_1 x}  ~ \text{and}~
 \phi^{3,4}(x)
=
\begin{pmatrix}
-V_{k_1} \\
0 \\
U_{k_1} \\
0
\end{pmatrix} e^{\pm i k_2 x},
\end{equation}
with
\begin{equation} U_{k_1} = \sqrt{\frac{\epsilon_k + g n_0 + E}{2 E}},  \quad
 V_{k_1} =  -\sqrt{\frac{\epsilon_k + g n_0 - E}{2 E}}. \end{equation}
The modes $\phi^{1,2}(x)$ with real $k$ are normalised physical solutions, corresponding to propagating phonons. $\phi^{3,4}(x)$ with imaginary $k$ are evanescent or growing waves, which
are unphysical but mathematically admissible.

The spectrum for the spin-down block from Eq.\eqref{bdg0} is
$
E = \pm(\epsilon_k + \hbar\Omega_0)$ with $\epsilon_k = {\hbar^2 k^2}/{2m},
$
giving four wave-numbers $ \pm k_3 $  and $ \pm k_4 $:
\begin{eqnarray}
k_3 &=& \begin{cases}
                 \sqrt{2m/\hbar^2} \sqrt{E - \hbar\Omega_0} \quad ~\text{for} \quad  E > \hbar\Omega_0 \\
                 i \sqrt{2m/\hbar^2}  \sqrt{\hbar\Omega_0 - E} \quad \text{for} \quad  E < \hbar\Omega_0,
               \end{cases}\\
k_4  &=& i \sqrt{2m/\hbar^2}  \sqrt{\hbar\Omega_0 + E},
\end{eqnarray}
and the corresponding eigenvectors are:
\begin{equation}
\phi^{5,6}(x)
= \begin{pmatrix} 0 \\ 1 \\ 0 \\0 \end{pmatrix}  e^{\pm i k_3 x} \quad \text{and} \quad   \phi^{7,8}(x)=
\begin{pmatrix}
0 \\
0 \\
0 \\
1
\end{pmatrix} e^{\pm i k_4 x}.
\end{equation}
Real $k$ describes propagating particles, and imaginary $k$ are for
evanescent or growing waves.

\subsection{Transfer  Matrix}

Since the BdG equations constitute a set of second-order differential equations, both the wavefunction $\phi(x)$ and its spatial derivative $\partial_x\phi(x)$ must be continuous across the entire space. To enforce these continuity conditions, we introduce the continuous augmented vector $
\mathbf{\Phi}(x) = \begin{pmatrix}
\phi(x) \\
\partial_x \phi(x)
\end{pmatrix}$.
In both the left and right regions outside the domain wall, the system is almost uniform and the wavefunction is a superposition of the eight modes. We define amplitude vectors
$
\mathbf{C}_L = (A_1,A_2,\cdots,A_8)^T$,
$\mathbf{C}_R = (A'_1,A'_2,\cdots,A'_8)^T$,
and the $8 \times 8$ matrix:
\begin{equation}
G(x) = \begin{pmatrix}
\phi^1(x) & \phi^2(x) & \cdots & \phi^8(x) \\
\partial_x \phi^1(x) & \partial_x \phi^2(x) & \cdots & \partial_x \phi^8(x)
\end{pmatrix},
\end{equation}
so that $\mathbf{\Phi}_L(x) =  G \mathbf{C}_L $ and $\mathbf{\Phi}_R(x) =  G \mathbf{C}_R$.

The BdG equation is a linear equation. Thus, wave propagation through the wall is described by a transfer matrix
$\mathcal{ M}$:
\begin{equation}
\mathbf{C}_R = \mathcal{ M} \, \mathbf{C}_L.
\end{equation}
To compute $\mathcal{ M}$ we rewrite the BdG equation as the linear ordinary differential equation (ODE)
\begin{equation}\label{ode1}
\partial_x \mathbf{\Phi}(x) = \mathbf{A}(x; E) \mathbf{\Phi}(x)
\end{equation}
with the $8 \times 8$ matrix
\begin{equation}
\mathbf{A}(x; E) =
\begin{pmatrix}
0 & 0 & \mathbb{I}_2 &  0  \\
0 & 0 &  0   & \mathbb{I}_2\\
\mathbf{K}_{UU} & \mathbf{K}_{UV}  & i \theta' \sigma_y & 0 \\
\mathbf{K}_{VU} & \mathbf{K}_{VV}  & 0 & i \theta' \sigma_y
\end{pmatrix}.
\end{equation}
The blocks are
\begin{eqnarray}
\mathbf{K}_{UU} &=& [\mathcal{B}(x) -E\mathbb{I}_2]/\alpha \\
\mathbf{K}_{VV} &=& [\mathcal{B}^*(x) + E\mathbb{I}_2]/\alpha, \\
\mathbf{K}_{UV} &=&  \mathbf{K}_{VU}^* = g \Psi_0 \Psi_0^T /\alpha,
\end{eqnarray}
with $\alpha = {\hbar^2}/{2m}$ and
\begin{eqnarray}
\mathcal{B}(x) &=& \frac{i\alpha}{2}\theta''(x)\sigma_y + \frac{\alpha}{4}[\theta'(x)]^2\mathbb{I}_2 - \frac{\hbar\Omega_0}{2}\sigma_z \notag\\
&& + [g n_0(x) - \mu]\mathbb{I}_2 + g \Psi_0 \Psi_0^\dagger.
\end{eqnarray}

Let $ x_L $ and $ x_R $ be the left and right boundary positions of the domain wall. We choose the eight left amplitude vector  $\mathbf{C}_L^{(j)}$ as $ A_{i}^{(j)} = \delta_{ij}$ for $ j = 1, \cdots, 8 $. Integrate the ODE \eqref{ode1} from $ x = x_L $ to $ x = x_R $ to obtain $\mathbf{\Phi}^{(j)}(x_R)$ and the corresponding right amplitude vector
\begin{equation}
\mathbf{C}_R^{(j)} = G^{-1}(x_R) \, \mathbf{\Phi}^{(j)}(x_R).
\end{equation}
Assembling these columns gives the transfer matrix
\begin{equation}
\mathcal{M} = \begin{pmatrix}
\mathbf{C}_R^{(1)} & \mathbf{C}_R^{(2)} & \cdots & \mathbf{C}_R^{(8)}
\end{pmatrix}
\end{equation}
which fully characterises the scattering properties.

\subsection{Scattering Coefficients}

\begin{figure}[tbp]
\centering
\includegraphics[width=\linewidth]{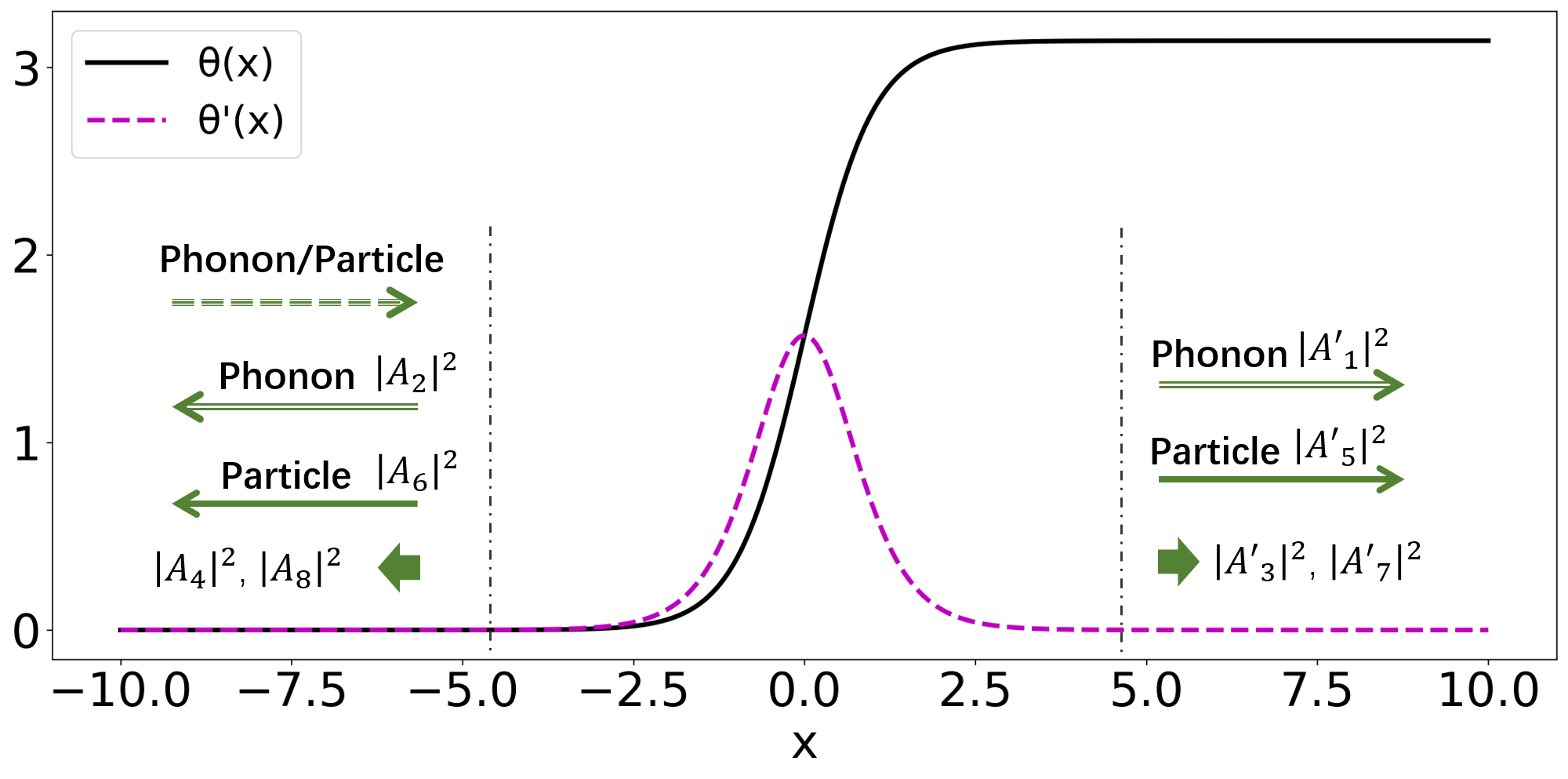}
\caption{Spatial profiles of the domain wall's spin rotation angle $\theta(x)$ and its derivative $\theta'(x)$ for a total twist $\Theta = \pi$, and schematic of the multi-channel scattering process for a linear wave incident from the left on the magnetic domain wall. An incident phonon (or particle) can be reflected as a phonon ($|A_2|^2$) or a particle ($|A_6|^2$), and transmitted as a phonon ($|A'_1|^2$) or a particle ($|A'_5|^2$). Evanescent waves ($|A_4|^2$, $|A_8|^2$ on the left and $|A'_3|^2$, $|A'_7|^2$ on the right) are also indicated. The scattering is governed by synthetic spin-orbit coupling induced by the domain wall texture and the mean-field background field.}
\label{fig:domain_wall}
\end{figure}

We now consider the scattering problem governed by
\begin{equation}\label{trm}
\mathbf{C}_R - \mathcal{M} \, \mathbf{C}_L = 0.
\end{equation}
For $E > \hbar\Omega_0$, we take a left-incident phonon (PN) (see Fig.~\ref{fig:domain_wall}).
The boundary conditions are
\begin{eqnarray}
&&A_1 = 1,\quad A_5 = 0,\quad A'_2 = 0,\quad A'_6 = 0, \label{condi1}\\
&&A_3 = 0,\quad A'_4 = 0,\quad A_7 = 0,\quad A'_8 = 0. \label{condi2}
\end{eqnarray}
The second line ensuring the wave-function remains finite at infinity. The unknown amplitudes are collected in
\begin{equation}
\bm{{C}} = (A'_1, A_2, A'_3, A_4, A'_5, A_6, A'_7, A_8)^T.
\end{equation}
From Eq.\eqref{trm}, we obtain
\begin{equation}\label{cdequ}
\mathcal{M}' \bm{{C}} = \bm D,
\end{equation}
where
\begin{equation}
\mathcal M' =
\begin{pmatrix}
1 & -M_{12} & 0 & -M_{14} & 0 & -M_{16} & 0 & -M_{18} \\
0 & -M_{22} & 0 & -M_{24} & 0 & -M_{26} & 0 & -M_{28} \\
0 & -M_{32} & 1 & -M_{34} & 0 & -M_{36} & 0 & -M_{38} \\
0 & -M_{42} & 0 & -M_{44} & 0 & -M_{46} & 0 & -M_{48} \\
0 & -M_{52} & 0 & -M_{54} & 1 & -M_{56} & 0 & -M_{58} \\
0 & -M_{62} & 0 & -M_{64} & 0 & -M_{66} & 0 & -M_{68} \\
0 & -M_{72} & 0 & -M_{74} & 0 & -M_{76} & 1 & -M_{78} \\
0 & -M_{82} & 0 & -M_{84} & 0 & -M_{86} & 0 & -M_{88}
\end{pmatrix},\end{equation}
and
\begin{equation}
{\bm D} = {\bm D}_{\text{PN}} =
(
M_{11},
M_{21},
M_{31},
M_{41},
\cdots,
M_{81}
)^T.
\end{equation}
Solving Eq.~\eqref{cdequ} yields the reflected probability amplitudes ($A_2$, $A_6$) and transmitted probability amplitudes ($A'_1$, $A'_5$) in the phonon and particle channels, respectively.

Next we consider a left-incident particle (PT). The boundary conditions become
\begin{eqnarray}
&&A_1 = 0,\quad A_5 = 1,\quad A'_2 = 0,\quad A'_6 = 0, \\
&&A_3 = 0,\quad A'_4 = 0,\quad A_7 = 0,\quad A'_8 = 0.
\end{eqnarray}
The same $\mathcal M'$ applies, but with
\begin{equation}
{\bm D} = {\bm D}_{\text{PT}} =
(
M_{15},
M_{25},
M_{35},
M_{45},
\cdots,
M_{85}
)^T.
\end{equation}
The resulting $\bm{{C}}$ yields
the probability amplitudes $(A_2, A_6)$ and $(A'_1, A'_5)$.

For $E < \hbar\Omega_0$, only the phonon branch propagates; the particle branch is evanescent. Using the same boundary conditions \eqref{condi1} [here $A_5 = 0$ and $A'_6 = 0$ because the wave function must be finite at infinity] and \eqref{condi2}, we solve Eq.\,\eqref{cdequ} to obtain the phonon reflected probability amplitude $A_2$ and transmitted probability amplitude  $A'_1$. With these coefficients, the scattering matrix also can be constructed.

\subsection{Numerical Calculation}

We now turn to the discrete model required for numerical work. Space is discretised on a uniform grid
\begin{equation}
x_i = i \Delta x, \quad \text{for } i = 0, 1, \dots, N-1,
\end{equation}
and the wave-function is denoted by $\chi_i \equiv \chi(x_i)= \begin{pmatrix}
U(x_i) \\
V(x_i)
\end{pmatrix} $ for the BdG equations \eqref{bdg2}. Starting from the continuous equations we obtain the discrete analogue
\begin{equation}\label{bdgd}{
\mathcal{K}_i \chi_{i+1} + \mathcal{N}_i \chi_i + \mathcal{L}_i \chi_{i-1} = 0.
}\end{equation}
with the forward and backward coupling matrices being
$
\mathcal{K}_i = K_i \mathbb{I}_2
$ and
$
\mathcal{L}_i =  L_i \mathbb{I}_2
$
where
\begin{eqnarray}
K_i &=& -\frac{\hbar^2}{2m(\Delta x)^2} \mathbb{I}_2 + i\frac{\hbar^2 [\theta'(x_{i+1})+ \theta'(x_i)]}{8m\Delta x} \sigma_y, \\
L_i &=& -\frac{\hbar^2}{2m(\Delta x)^2} \mathbb{I}_2 - i\frac{\hbar^2 [\theta'(x_{i})+\theta'(x_{i-1})]}{8m\Delta x} \sigma_y,
\end{eqnarray}
and the on-site matrix
$
\mathcal{N}_i = \begin{pmatrix}
{N}_{i} - E\mathbb{I}_2 & {N}_{i\times} \\
{N}_{i\times}^* & {N}_{i}^* +  E\mathbb{I}_2
\end{pmatrix}
$
with  ${N}_{i\times}  = g(\Psi_0\Psi_0^T)_i $ and
\begin{eqnarray}
{N}_i &=& \left\{\frac{\hbar^2}{m(\Delta x)^2}  + \frac{\hbar^2 \left[{\theta'(x_i)}\right]^2}{8m}
 + g n_0(x_i) \right\} \mathbb{I}_2 \notag\\
 && - \frac{\hbar\Omega_0}{2}\sigma_z - \mu\mathbb{I}_2 + g(\Psi_0\Psi_0^\dagger)_i.
\end{eqnarray}
The resulting transfer-matrix form is
\begin{equation} \label{ode}
\begin{pmatrix} \chi_i \\ \chi_{i+1} \end{pmatrix} =
\begin{pmatrix}
0 & \mathbb{I}_4 \\
-\mathcal{K}_i^{-1} \mathcal{L}_i & -\mathcal{K}_i^{-1} \mathcal{N}_i
\end{pmatrix}
\begin{pmatrix} \chi_{i-1} \\ \chi_i \end{pmatrix}.
\end{equation}

Away from the domain wall where $\theta'(x) =0$ the eigenvector retains the continuum form $\phi^j(x_i)$ but the wave-vectors (denotes $\tilde k$) are corrected according to
\begin{equation}
  \epsilon_k = \frac{\hbar^2 k^2}{2m} = \frac{\hbar^2}{2m} \frac{4}{\Delta x^2} \sin^2 \left(\frac{\tilde k \Delta x}{2}\right),
\end{equation}
giving
$
\sin({\tilde{k}_{\lambda} \Delta x}/{2} ) = {{k}_{\lambda} \Delta x}/{2}$,
for $\lambda =1,\cdots,4$.

The general wave-function outside the wall is the superposition of
$ \begin{pmatrix}
\phi^j(x_i) \\
\phi^j(x_{i+1})
\end{pmatrix}
$.
The relation between the left amplitude vector $
\mathbf{C}_L $
and the right amplitude vector $
\mathbf{C}_R $  is again
$\mathbf{C}_R = \mathcal{ M} \, \mathbf{C}_L$.
We set  $\mathbf{C}_L^{(j)}$ as $ A_{i}^{(j)} = \delta_{ij}$ for $ j = 1, \cdots, 8 $, propagate $ \mathbf{\Phi}^{(j)}(x)$  from $x_L$ to $x_R$ with the transfer equation, and extract
\begin{equation}
\mathbf{C}_R^{(j)} = {\tilde G}^{-1} \, \mathbf{\Phi}^{(j)}(x_R).
\end{equation}
where
\begin{equation}
\tilde{G} = \begin{pmatrix}
\phi^1(x_R) & \phi^2(x_R) & \cdots & \phi^8(x_R) \\
\phi^1(x_R+\Delta x) & \phi^2(x_R+\Delta x) & \cdots &  \phi^8(x_R+\Delta x)
\end{pmatrix}.
\end{equation}
Assembling the columns gives the transfer matrix in the discrete model:
 $\mathcal M = (
\mathbf{C}_R^{(1)}, ~ \mathbf{C}_R^{(2)}, ~ \cdots ~ \mathbf{C}_R^{(8)}
)$, from which all scattering coefficients can be obtained.

\subsection{Conserving Current}
To obtain the conserved current it is convenient to recast the BdG equation \eqref{bdg2}  in the form
\begin{equation}\label{bdg3}
\mathcal{H}'_{\text{BdG}} \begin{pmatrix} U(x) \\ V(x) \end{pmatrix} = E \begin{pmatrix} U(x) \\ -V(x) \end{pmatrix}.
\end{equation}
where $\mathcal{H}'_{\text{BdG}} = \mathcal{T} + \mathcal{V} $. Left-multiplying Eq.\eqref{bdg3} by $\tilde{\Psi}^\dagger = (U^\dagger, V^\dagger )$, gives
$
\tilde{\Psi}^\dagger \mathcal{H}'_{\text{BdG}}  \tilde{\Psi} = E \, \tilde{\Psi}^\dagger \tau_z \tilde{\Psi}$.
Subtracting its Hermitian conjugate yields
$
\tilde{\Psi}^\dagger \mathcal{H}'_{\text{BdG}}  \tilde{\Psi} - ( \tilde{\Psi}^\dagger \mathcal{H}'_{\text{BdG}} \tilde{\Psi} )^\dagger = 0$.
Using the Hermiticity of $\mathcal{V}$, the left-hand side becomes
$
\tilde{\Psi}^\dagger \mathcal{T} \tilde{\Psi} - ( \mathcal{T} \tilde{\Psi} )^\dagger \tilde{\Psi} = 0$,
leading to a total derivative equation $
 \partial_x [ \tilde{\Psi}^\dagger D_x \tilde{\Psi} - ( D_x \tilde{\Psi} )^\dagger \tilde{\Psi} ] =\tilde{\Psi}^\dagger D_x^2 \tilde{\Psi} - ( D_x^2 \tilde{\Psi} )^\dagger \tilde{\Psi}  = 0$.
The conserved current is therefore
\begin{eqnarray}
&J(x)& = \frac{\hbar}{2mi} \left[ \tilde{\Psi}^\dagger D_x \tilde{\Psi} - ( D_x \tilde{\Psi} )^\dagger \tilde{\Psi} \right] \\
&=& \frac{\hbar}{2mi}  \big[ U^\dagger D_x U - \left( D_x U \right)^\dagger U + V^\dagger D_x V - \left( D_x V \right)^\dagger V \big]. \notag
\end{eqnarray}
This requires the currents at left and right sides should be equal, $J_L = J_R$.

For a left-incident phonon with $E > \hbar \Omega_0$ the currents on the left and right are
\begin{eqnarray}
J_L &=&  (U_{k_1}^2 + V_{k_1}^2) (1 - |A_2|^2) \frac{\hbar k_1}{m} - |A_6|^2 \frac{\hbar k_3}{m}, \\
J_R &=& |A'_1|^2 (U_{k_1}^2 + V_{k_1}^2)\frac{\hbar k_1}{m}  + |A'_5|^2 \frac{\hbar k_3}{m}. \end{eqnarray}
The incident current is $j_i = (U_{k_1}^2 + V_{k_1}^2) {\hbar k_1}/{m}$.
The phonon reflection and transmission coefficients are given by the corresponding current over the incident current, resulting
\begin{equation} \label{rtrt}
  R_{\text{PN}} = |A_2|^2, \quad \text{and}\quad T_{\text{PN}} = |A'_1|^2,
\end{equation}
respectively. Moreover, both left-moving and right-moving particles are generated as a result of the collision with the domain wall. The corresponding particle reflection and transmission coefficients are
\begin{equation}
  R_{\text{PT}} = \frac{|A_6|^2 k_3}{k_1  (U_{k_1}^2 + V_{k_1}^2)} , \quad T_{\text{PT}} = \frac{|A'_5|^2  k_3}{k_1  (U_{k_1}^2 + V_{k_1}^2)}.  \\
\end{equation}
$J_L = J_R$ yields the probability current conservation relation \cite{ohashi2021perfect},
\begin{equation} F \equiv T_{\text{PN}} +  R_{\text{PN}} + T_{\text{PT}} + R_{\text{PT}}  =1. \end{equation}
For $E< \hbar \Omega_0$, only the phonon branch propagates and
\begin{equation}  F \equiv  T_{\text{PN}} +  R_{\text{PN}} =1, \end{equation}
where the phonon reflection and transmission coefficients are the same as those in Eq.~\eqref{rtrt}.

For a left-incident particle with  $E> \hbar \Omega_0$, one finds
\begin{eqnarray} J_L &=&  \frac{\hbar k_3}{m}  - |A_2|^2 (U_{k_1}^2 + V_{k_1}^2) \frac{\hbar k_1}{m} - |A_6|^2 \frac{\hbar k_3}{m}, \\
J_R &=& |A'_1|^2 (U_{k_1}^2 + V_{k_1}^2)\frac{\hbar k_1}{m}  + |A'_5|^2 \frac{\hbar k_3}{m}. \end{eqnarray}
The incident current is $j_i =  {\hbar k_3}/{m}$. The phonon reflection and transmission coefficients are
\begin{equation}
  R_{\text{PN}} = |A_2|^2 \frac{k_1  (U_{k_1}^2 + V_{k_1}^2)}{k_3}, \quad
   T_{\text{PN}} = |A'_1|^2 \frac{k_1  (U_{k_1}^2 + V_{k_1}^2)}{k_3}.
\end{equation}
The particle reflection and transmission coefficients are
\begin{equation}
  R_{\text{PT}} = |A_6|^2 , \quad T_{\text{PT}} = |A'_5|^2.  \\
\end{equation}
$J_L = J_R$ leads to
\begin{equation}
 F \equiv   T_{\text{PN}} +  R_{\text{PN}} + T_{\text{PT}} + R_{\text{PT}}  =1. \end{equation}
The relation $F=1$ guarantees conservation of the quasiparticle/particle probability current and provides important consistency checks for the calculated reflection and transmission coefficients.

\subsection{Results}

\begin{figure}[h!]
\centering
\includegraphics[width=\linewidth]{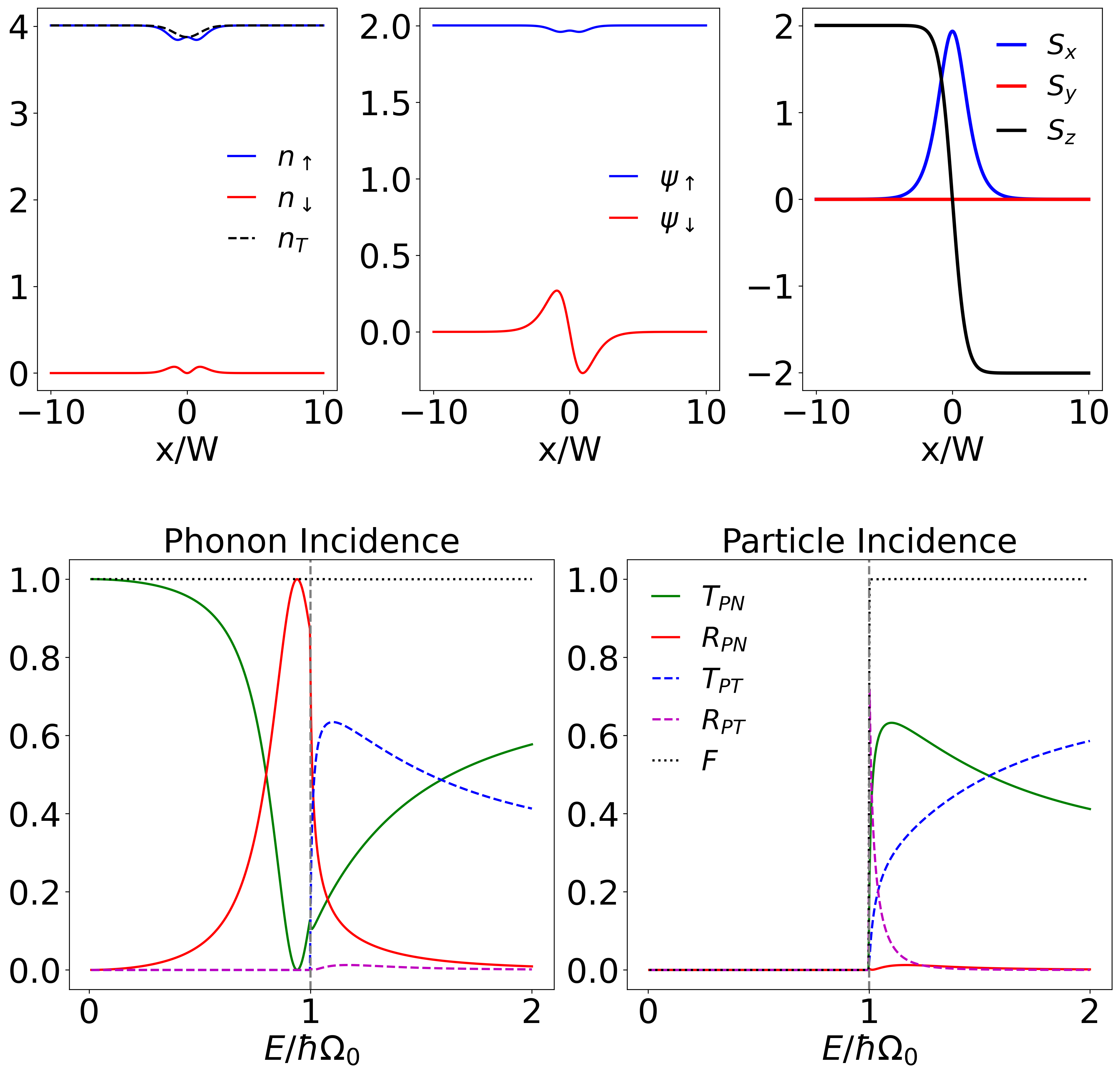}
\caption{(Top) Ground-state properties for a $\Theta=\pi$ domain wall: BEC density profile, wavefunction components (imaginary parts vanish), and spin texture. (Bottom) Energy-dependent scattering coefficients for phonon/particle incidence with $E \in [0.01,2]\hbar\Omega_0$. The scattering coefficients are defined as follows: $T_{\text{PN}}$ (phonon transmission), $R_{\text{PN}}$ (phonon reflection), $T_{\text{PT}}$ (particle transmission), and $R_{\text{PT}}$ (particle reflection). Current conservation is verified via $F=1$. The ground state is obtained for a system of length $L=20$ with total particle number $N=80$, discretized on a uniform grid with spacing $\Delta x = 1/30$. The unit for density $n$ is $1/W$, and the unit for the wavefunction is $1/\sqrt{W}$.}
\label{fig10}
\end{figure}

\begin{figure}[h!]
\centering
\includegraphics[width=\linewidth]{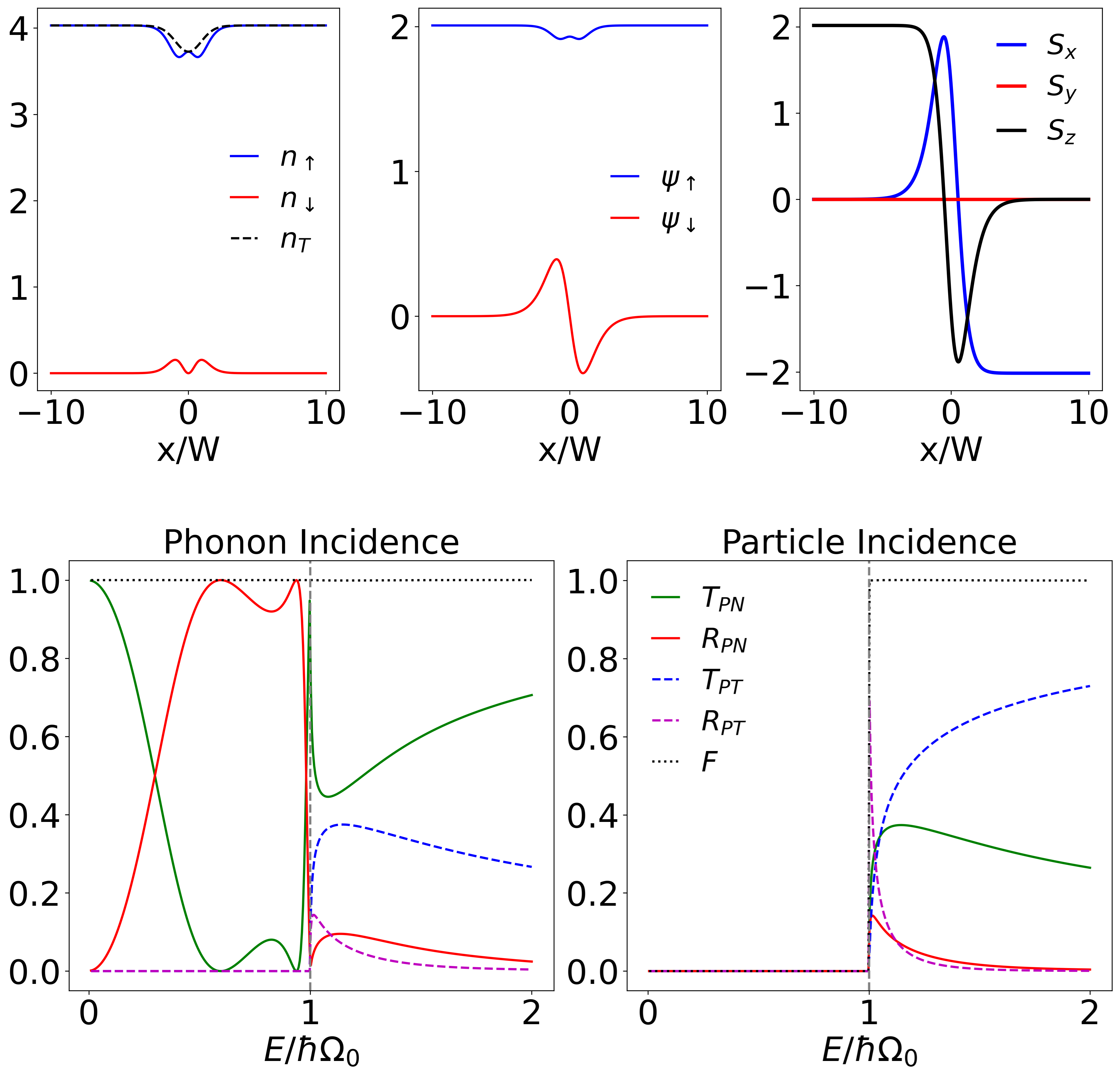}
\caption{Ground-state profile and scattering coefficients for linear waves incident on a domain wall with twist $\Theta = 3\pi/2$. The unit for density $n$ is $1/W$, and the unit for the wavefunction is $1/\sqrt{W}$.}
\label{fig15}
\end{figure}

\begin{figure}[h!]
\centering
\includegraphics[width=\linewidth]{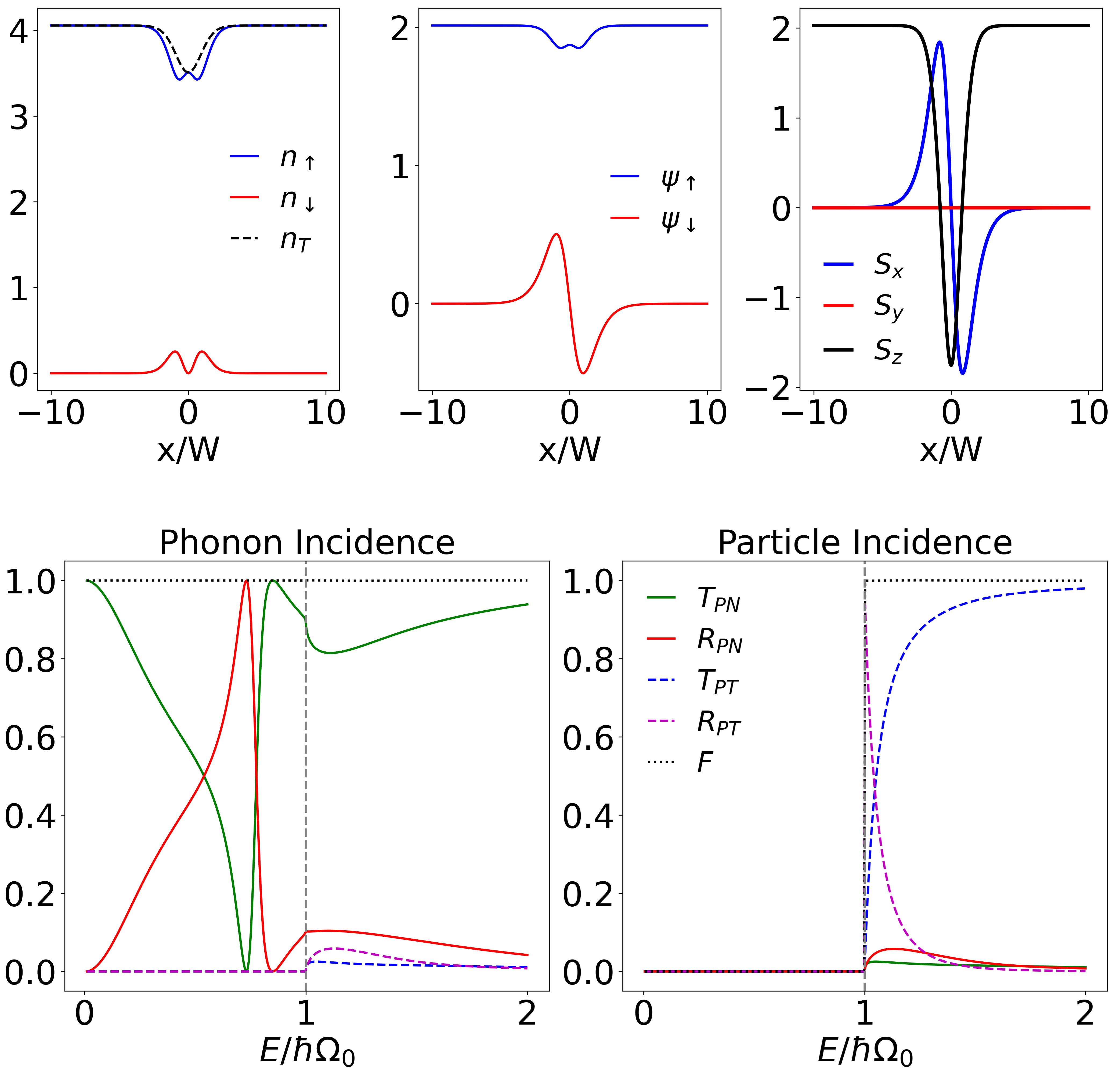}
\caption{Ground-state profile and scattering coefficients for linear waves incident on a domain wall with twist $\Theta = 2\pi$. The unit for density $n$ is $1/W$, and the unit for the wavefunction is $1/\sqrt{W}$.}
\label{fig20}
\end{figure}

\begin{figure}[t!]
\centering
\includegraphics[width=\linewidth]{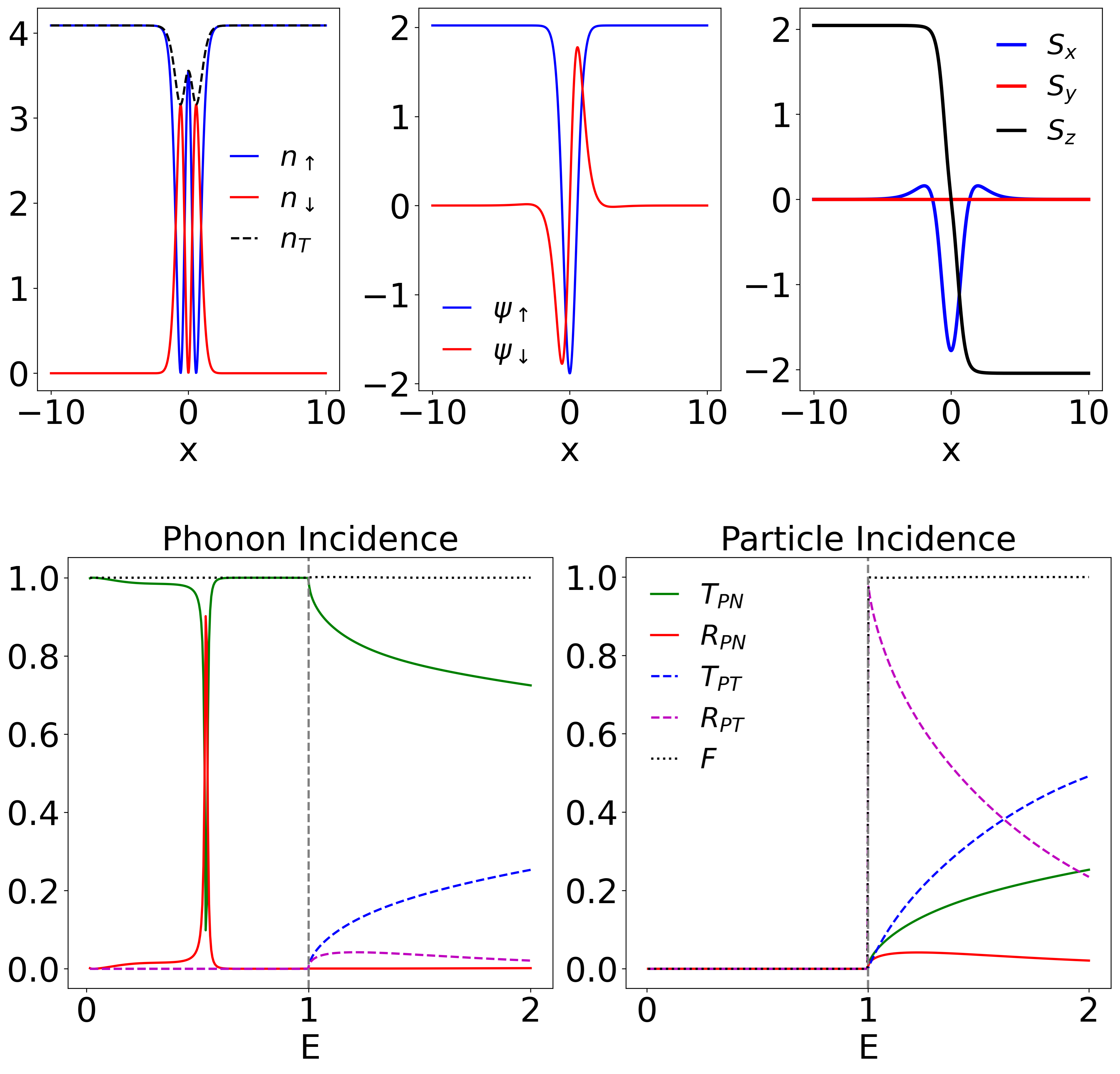}
\caption{Ground-state profile and scattering coefficients for linear waves incident on a domain wall with twist $\Theta = 3\pi$. The unit for density $n$ is $1/W$, and the unit for the wavefunction is $1/\sqrt{W}$.}
\label{fig30}
\end{figure}

\begin{figure}[t!]
\centering
\includegraphics[width=\linewidth]{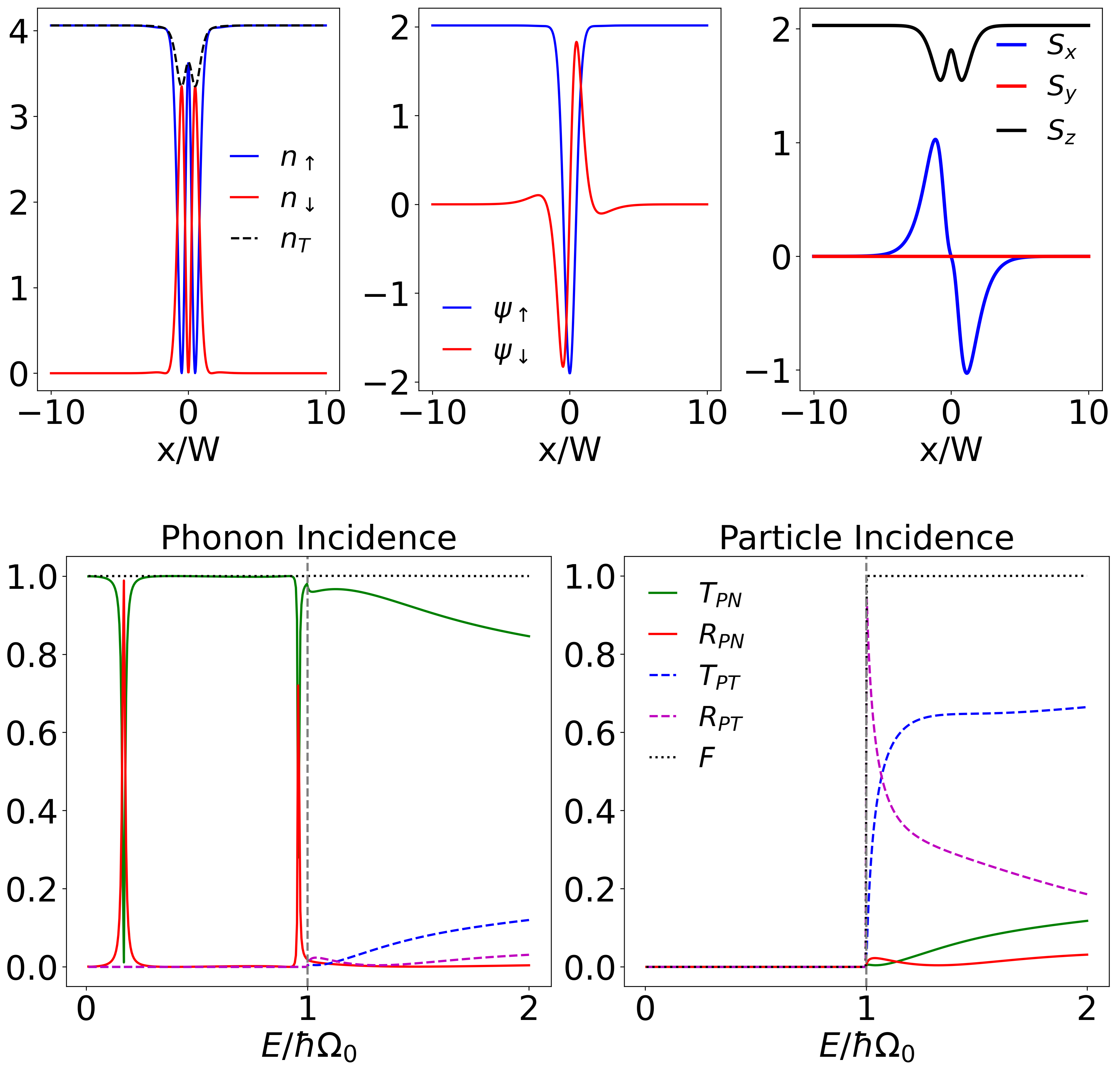}
\caption{Ground-state profile and scattering coefficients for linear waves incident on a domain wall with twist $\Theta = 4\pi$. The unit for density $n$ is $1/W$, and the unit for the wavefunction is $1/\sqrt{W}$.}
\label{fig40}
\end{figure}

\begin{figure}[t!]
\centering
\includegraphics[width=\linewidth]{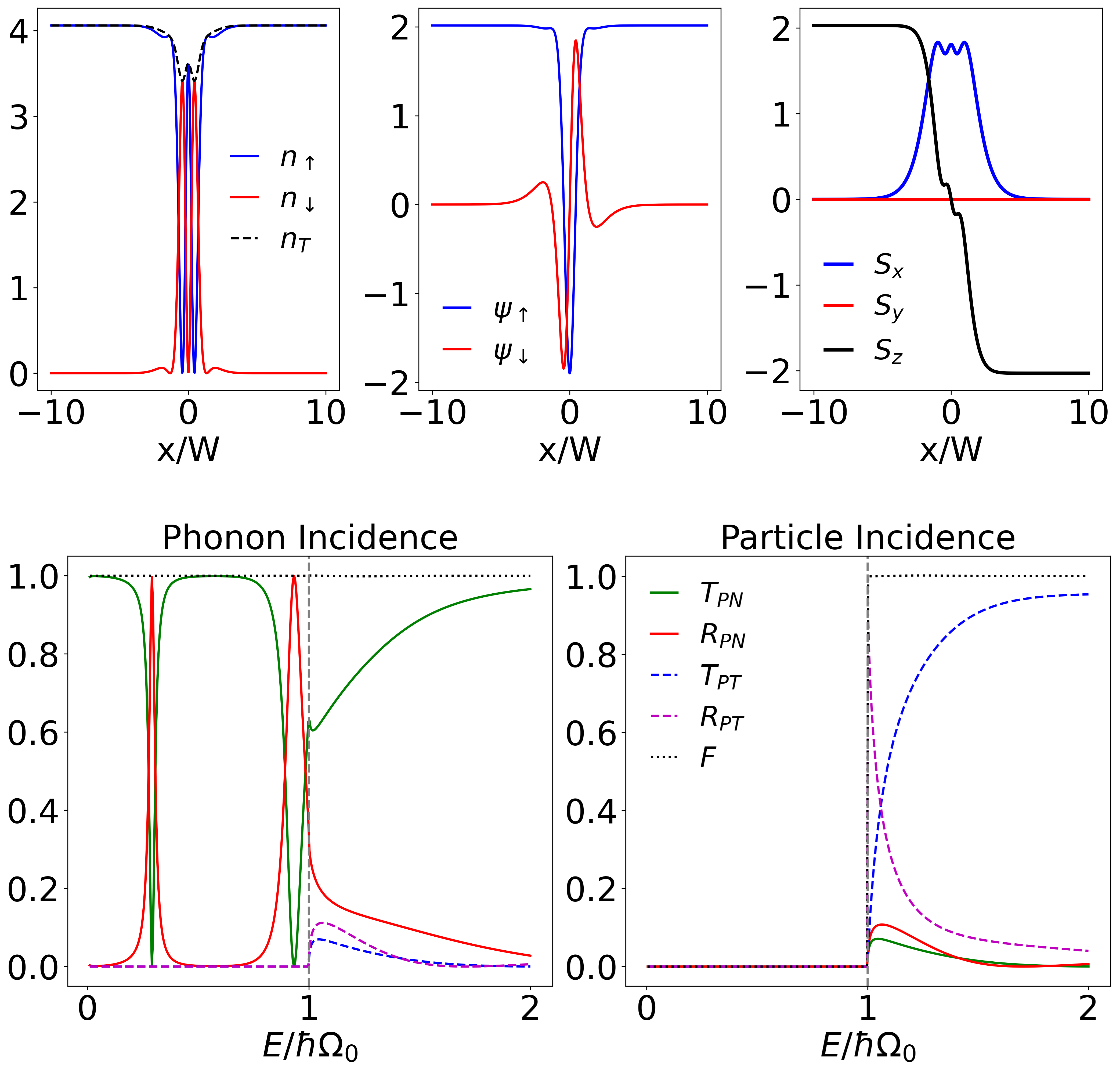}
\caption{Ground-state profile and scattering coefficients for linear waves incident on a domain wall with twist $\Theta = 5\pi$. The unit for density $n$ is $1/W$, and the unit for the wavefunction is $1/\sqrt{W}$.}
\label{fig50}
\end{figure}

We now compute the scattering coefficients. The results for systems with different twist angles $\Theta$ are presented in Figs.~\ref{fig10}-\ref{fig50}. All calculations are performed in natural units with $\hbar = m = 1$, the Zeeman splitting energy taken as the energy unit ($\hbar\Omega_0 = 1$), and the domain wall width as the length unit ($W = 1$). The interaction strength is fixed at $g = 1$ (the effective interaction magnitude can be tuned via the particle number). The transport properties are evaluated for energies $E$ in the range $[0.01, 2]$ (in units of $\hbar\Omega_0$).

Numerical analysis reveals distinct scattering behaviors that depend critically on both the twist angle $\Theta$ and the excitation energy $E$. The results exhibit pronounced energy dependence with a clear threshold at $E = \hbar\Omega_0$. Below this threshold, scattering occurs exclusively through phonon channels. Above $E = \hbar\Omega_0$, both phonon and particle channels become accessible, leading to multi-channel scattering phenomena. Current conservation checks confirm the reliability of our transfer matrix approach ($F=1$).


Figure \ref{fig10} presents the results for a domain wall with a full $\pi$ twist. In the rotated spin frame, the ground-state wavefunction remains close to $(\sqrt{n_0}, 0)^T$, indicating that the spin orientation does not deviate significantly from the local Zeeman field direction. A noticeable deviation occurs within the domain wall region, primarily due to the spin-orbit coupling in the gauge field. This is further corroborated by calculating the spin expectation values $S_i = \Psi_{\text{lab},0}^\dagger (\sigma_i/2) \Psi_{\text{lab},0}$ ($i=x,y,z$) in the original laboratory frame, which demonstrates a complete $\pi$ rotation of the spin texture across the wall in the $S_z$-$S_x$ plane. The lower panels show the scattering coefficients. For energies below the Zeeman threshold, scattering occurs solely through the phonon channel. At small twist angles, phonons transmit almost transparently. As $\Theta$ increases from $0$ to $\pi$, a resonance feature emerges near $E = \hbar\Omega_0$. Above the threshold, the particle channel opens, leading to a finite transmission coefficient into the particle mode, $T_{\text{PT}}$. Figure \ref{fig10} (Bottom) shows the energy-dependent scattering coefficients for phonon and particle incidence in the range of $E \in [0.01, 2]\hbar\Omega_0$. In the phonon-incidence case (left panel), the phonon transmission coefficient  $T_{\text{PN}}$ approaches unity in the low-energy limit and decreases monotonically with increasing energy, while the phonon reflection coefficients $R_{\text{PN}}$ rises correspondingly from near zero. This confirms the anomalous tunneling phenomenon, in which elementary excitations in a BEC transmit through a potential barrier with probability one in the low-energy limit \cite{Kovrizhin2001,Kagan2003,Danshita2006,Tsuchiya2008,Watabe2008,Watabe2012,Watabe2015,Watabe2011,Watabe2011a,Watabe2011b,Kato2008}. This effect has been observed in both one-component and multi-component BECs across various types of barriers. In the zero-energy limit, the excitations correspond to the Nambu-Goldstone mode and become identical to the condensate particles themselves. Consequently, their wave function extends across both sides of the barrier, leading to perfect transmission. In spin-1 system, besides the anomalous tunneling mode, there is a transverse mode with perfect reflection in the low-energy limit \cite{Watabe2012}. For $E > \hbar\Omega_0$, the particle transmission coefficient $T_{\text{PT}}$ exhibits a nonmonotonic behavior and is comparable to the phonon transmission coefficient, while the particle reflection coefficient $R_{\text{PT}}$ remains close to zero throughout the entire energy range. In the case of particle incidence (right panel), the particle reflection coefficient $R_{\text{PT}}$ is relatively large when $E \approx \hbar\Omega_0$, and it quickly decreases with energy. For energies $E > \hbar\Omega_0$, the phonon transmission coefficient $T_{\text{PN}}$ exhibits nonmonotonic behavior and is comparable to the particle transmission coefficient, indicating that incident particles can excite phonon modes at higher energies. Moreover, the phonon reflection coefficient remains near zero throughout the entire energy range.

Figure \ref{fig15} shows results for $\Theta = 3\pi/2$. While the ground-state wavefunction in the rotated spin frame appears qualitatively similar to the $\Theta = \pi$ case, the spin expectation values in the original frame reveal a complete $3\pi/2$ rotation of the spin orientation across the domain wall. This altered spin texture leads to significantly different phonon excitations and thus scattering behavior compared to the $\Theta = \pi$ case. For energies below the Zeeman threshold, the phonon scattering coefficient is notably modified, except for the anomalous tunneling phenomenon in the low-energy limit, as shown in the bottom panel of Figure \ref{fig15}. For the phonon-incidence case with energies $E > \hbar\Omega_0$, the phonon transmission coefficient $T_{\text{PN}}$ is significantly larger than the particle transmission coefficient $T_{\text{PT}}$, indicating that incident phonons become harder to transition into particles compared to the $\Theta = \pi$ case. A similar phenomenon occurs for the particle-incidence case, where the particle transmission is mostly larger than the phonon transmission. We conclude that, above the threshold, the transition between collective (phonon) and single-particle excitations is suppressed in the $\Theta = 3\pi/2$ case compared to the $\Theta = \pi$ case.

Figure \ref{fig20} shows the results for a twist angle $\Theta = 2\pi$. While the ground-state wavefunction in the rotated spin frame is qualitatively similar to previous cases, the spin texture in the original laboratory frame completes a full $2\pi$ rotation, resulting in identical spin orientations on both sides of the domain wall. This restored symmetry leads to distinct scattering characteristics. In particular, for energies above the threshold, processes where a phonon (particle) transitions into a particle (phonon) correspond to a spin flip in the original spin space. These spin-flip transitions are strongly suppressed during the scattering process. From the scattering properties shown at the bottom of Figure \ref{fig20}, it is evident that in the case of phonon incidence (left panel), when the energy exceeds the Zeeman threshold, both the particle transmission and reflection coefficients are notably smaller than those in the previously considered twist-angle configurations. In contrast, for particle incidence above the Zeeman threshold, both the phonon transmission and reflection coefficients become rather small. We conclude that, when the spin polarization is the same on both sides in the original frame, the transition between collective (phonon) and single-particle excitations is nearly forbidden throughout the entire energy range.

Figure \ref{fig30} displays results for a twist angle $\Theta = 3\pi$. Remarkably, the effective spin rotation (in the $S_z$-$S_x$ plane) in the original laboratory frame is $-\pi$ rather than $3\pi$. The total spin rotation is not consistent with the twist of the Zeeman field, reflecting a competition among Zeeman, kinetic, and interaction energies. The system becomes nearly transparent for excitations below the Zeeman threshold ($E < \hbar\Omega_0$).  However, the mismatch between the spin twist and the Zeeman field orientation generates a comb-like density modulation within the domain wall. In the case of phonon incidence, this structure induces pronounced Fano-like resonances below threshold \cite{Miroshnichenko2010,Vicencio2007,yasir2016controlled,Zheng2018}, visible as sharp dips in the phonon transmission coefficient $T_{\text{PN}}$ and corresponding peaks in the phonon reflection coefficient $R_{\text{PN}}$. Similar resonances have been observed in optical lattice systems combined with potential barriers, which possess nonuniform comb-like density structures \cite{Nakayama2015}. In the energy region above the Zeeman threshold, it opens additional scattering channels, which may account for the absence of Fano resonances in this regime. However, the transition between collective (phonon) and single-particle excitations is suppressed in the $\Theta = 3\pi$ case compared to the $\Theta = \pi$ case.   For particle incidence, the reflection coefficient $R_{\text{PT}}$ becomes comparable to the transmission coefficient $T_{\text{PT}}$. The particle becomes harder to pass through the domain wall in the comb-like structure.

Figure \ref{fig40} presents the intriguing case of $\Theta = 4\pi$. Here, the ground-state wavefunction in the rotated spin frame differs significantly from the $\Theta = 2\pi$ case. The effective spin rotation in the $S_z$-$S_x$ plane is $0$ rather than $4\pi$. With identical spin orientation on both sides, the system becomes nearly transparent for excitations below the Zeeman threshold ($E < \hbar\Omega_0$). Similar to the previous case, the mismatch between the spin and the Zeeman field twist leads to a comb-like density modulation, inducing two Fano-like resonances below threshold. In the energy region above the Zeeman threshold, the transition between collective (phonon) and single-particle excitations is further suppressed compared to the $\Theta = 3\pi$ case. For particle incidence, the reflection coefficient $R_{\text{PT}}$ decreases with energy and the transmission coefficient $T_{\text{PT}}$ increases with energy, like the behavior shown in Figure \ref{fig30}.


Figure \ref{fig50} shows the results for a twist angle $\Theta = 5\pi$. The ground-state configuration in the rotated spin frame differs from the $\Theta = \pi$ case but is similar to the $\Theta = 3\pi$ cases. The effective spin rotation in the original spin frame reduces to $\pi$ rather than $5\pi$ for the ground state. Similar to the $\Theta = 3\pi, 4\pi$ cases, the comb-like density gives rise to Fano-like resonances below the Zeeman threshold while simultaneously strongly suppressing transitions between collective (phonon) and single-particle excitation channels above threshold. From the bottom panels of Figure \ref{fig50}, it can be seen that in the case of phonon incidence (left panel), when the energy is below the Zeeman threshold, the variations of the phonon reflection and transmission coefficients are similar to those for the twist angle $4\pi$. Moreover, in all figures, when the energy exceeds the Zeeman threshold, the particle transmission and reflection coefficients for the phonon-incidence case are similar to the phonon transmission and reflection coefficients for the particle-incidence case. This property indicates transition between phonon and particle is mutual.



\begin{figure}[t!]
\centering
\includegraphics[width=\linewidth]{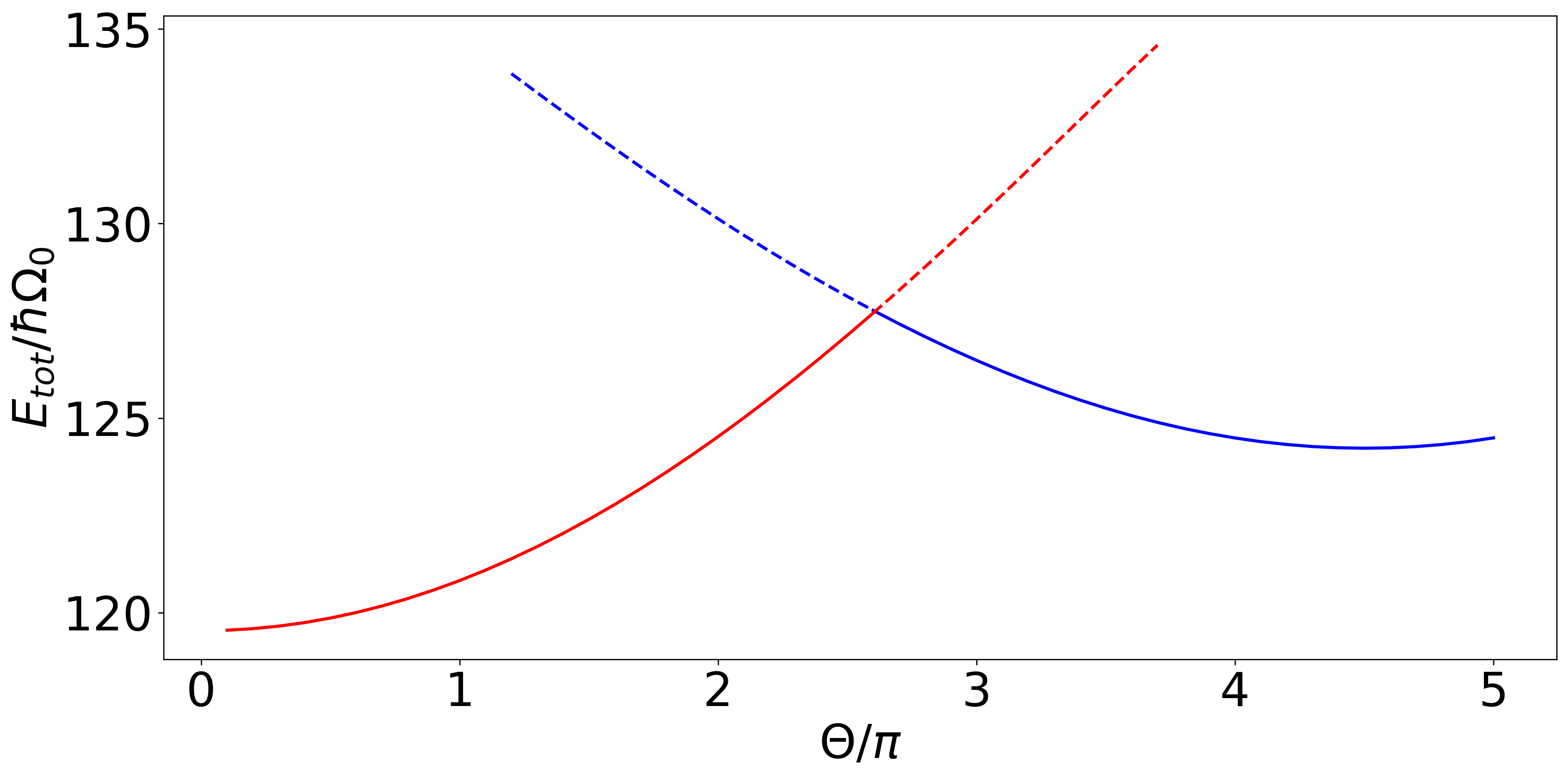}
\caption{Ground-state energy as a function of $\Theta$. The red lines represent the solution where the spin twist agrees with the Zeeman field twist, while the blue lines correspond to the solution where it does not. }
\label{fig8}
\end{figure}

We now comment on the comb-like density profile. Indeed, starting from different initial states, the GP equation can yield different stationary solutions around the critical value $\Theta_c \simeq 2.7\pi$. We compare the total energy to determine the true ground state. In Figure \ref{fig8}, we present the total energy for the ground state (solid lines) and metastable states (dashed lines). The red lines represent the solution where the spin twist agrees with the Zeeman field twist, while the blue lines correspond to the solution where it does not.

\section{Conclusion}

This work has established a comprehensive theoretical framework for understanding quantum scattering phenomena induced by magnetic domain walls in spin-1/2 BECs. Through systematic numerical investigation of varying twist angles $\Theta$, we have uncovered fundamental principles governing matter-wave transport in these geometrically tunable systems.

Our analysis reveals a sharp scattering threshold at the Zeeman energy $E = \hbar\Omega_0$, which cleanly separates regimes of pure phonon transport from multi-channel scattering involving both collective and single-particle excitations. A particularly striking finding is the critical twist angles ($\Theta_c \simeq 2.7\pi$), where the effective spin rotation in the laboratory frame reduces to values much smaller than the imposed $\Theta$. This geometric frustration generates strain fields that fundamentally reshape scattering properties, producing comb-like density modulations and pronounced Fano-like resonances below threshold $\hbar\Omega_0$. Above the Zeeman energy, the transition between phonon and particle are
suppressed in the comb-like structure.

The symmetry of boundary spin orientations emerges as a crucial control parameter for $\Theta<\Theta_c$. Systems with identical spin orientations on both sides of the domain wall exhibit enhanced transparency and strongly suppressed inter-channel transitions above the threshold $\hbar\Omega_0$. In contrast, configurations with opposite boundary orientations facilitate significant mode conversion between phonon and particle channels. This symmetry dependence makes the twist angle $\Theta$ a powerful and continuous parameter for engineering scattering pathways.

Methodologically, we have developed a unified transfer-matrix approach within the BdG framework that consistently treats all scattering channels while maintaining exact current conservation. This formalism, combined with high-precision ground-state determination, provides a robust computational tool for studying quantum transport in spatially inhomogeneous condensates. The phenomena predicted here are experimentally accessible in current ultracold-atom platforms using $^{87}\mathrm{Rb}$ or $^{23}\mathrm{Na}$ condensates with synthetic spin-orbit coupling.


In summary, magnetic domain walls with controlled twist angles constitute a versatile and highly tunable platform for manipulating matter-wave transport. By elucidating the intricate interplay between twist phase, mean-field interactions, and Zeeman coupling, we have demonstrated precise control over scattering resonances and mode transitions. These findings open new avenues for fundamental exploration of quantum many-body physics and technological applications in quantum simulation and atomtronics, where engineered scattering elements could enable novel matter-wave circuits and sensors.
\begin{acknowledgments}
This work is supported by the National Natural Science Foundation of China under grants No.\,12575028, No.\,12105223, No.\,12575029, No.\,12175180, and No.\,12247103, and Natural Science Basic Research Program of Shaanxi under Grant No.\,2024JC-YBMS-022.
\end{acknowledgments}

\end{document}